\documentclass[preprint,aps,showpacs]{revtex4}

\usepackage{graphicx}
\usepackage{dcolumn}
\usepackage{bm}

\begin{document}


\title{Growing length scale in gravity-driven dense granular flow}

\author{Shubha Tewari}
\email{stewari@mtholyoke.edu}
\author{Bidita Tithi}
 \affiliation{Department of Physics, Mount Holyoke College, 50 College Street, South Hadley, MA 01075}
\author{Allison Ferguson}%
\affiliation{Department of Biochemistry, University of Toronto, Toronto, Ontario M5S 1A8}

\author{Bulbul Chakraborty}
\email{bulbul@brandeis.edu}
\affiliation{Martin Fisher School of Physics, Brandeis University,
Mailstop 057, Waltham, MA 02454-9110 }

\date{\today}

\begin{abstract}
We report simulations of a two-dimensional, dense, bidisperse system
of inelastic hard disks falling down a vertical tube under the
influence of gravity. We examine the approach to jamming as the
average flow of particles down the tube is slowed by making the
outlet narrower. Defining coarse-grained velocity and stress fields,
we study two-point temporal and spatial correlation functions of
these fields in a region of the tube where the time-averaged
velocity is spatially uniform. We find that fluctuations in both
velocity and stress become increasingly correlated as the system
approaches jamming. We extract a growing length scale and time scale
from these correlations.
\end{abstract}

\maketitle

\section{\label{sec:level1}Introduction}
Granular materials (powders, seeds, grains, sand) consist of
macroscopic particles that interact via dissipative short-ranged or
contact forces \cite{jnb_rmp}. Granular systems are athermal, since
the characteristic energies needed to move a single grain are many
orders of magnitude larger than thermal energies. In the absence of
external forces, there is no motion; and the state of the system
under the application of external forces varies with the magnitude
of the force as well as with the packing density of the grains. In
this paper, we will describe results from numerical simulations of
the dense gravity-driven flow of grains down a vertical tube, as in
an hourglass, and the transition from a flowing state to one that is
stuck, or jammed.

The questions addressed here fall within the rubric of a proposal
\cite{liunagel} to unify disparate systems under a common framework,
that posits that there are some universal aspects to the slowing
down of the dynamics in many disordered systems as they move from a
mobile state to one that is frozen; such a transition is labeled a
jamming transition. There is no static structural signature of the
transition from a mobile to a jammed state: unlike first order
freezing transitions, there is no discontinuous change in the
density or broken translational symmetry. Nor is there a clear
signature of a diverging length scale derived from a two-point
static correlation function as in a second-order phase transition,
though there are other indications of a diverging length scale as a
critical density is approached \cite{silbert, wyart, teitel}.

A dense column of grains in a vertical hopper flows down at a steady
rate rather than accelerating under gravity because the weight of
the column is supported by the walls. The rate of flow decreases as
the size of the opening at the outlet of the hopper is decreased,
and ultimately jams when the opening is a few particle diameters
across. It is well known that the distribution of load in a static
column of sand is spatially inhomogeneous and organized along linear
structures called force chains \cite{dantu,chliu}. The question
remains open as to whether these structures begin to form in the
flowing state as the flow slows. In this article, we present
evidence from simulations for increasing spatial correlations in
both velocity and stress fluctuations as the flow rate decreases. We
extract a length scale from these correlations, and find that the
flow rate dependence of the length scales for velocity and stress
are in exact correspondence. Our results agree very well with recent
experimental observations \cite{gardel} of growing spatial
correlations in velocity fluctuations as the flow in a vertical
hopper approaches the jamming threshold, and help clarify how these
correlations arise in a flow that is dense, continuous, and highly
collisional.

There have been many efforts towards extracting a length scale in
granular systems. Inhomogeneous force chains were visualized in
sheared systems using photoelastic beads \cite{howell} and their
spatial correlations quantified \cite{majmudar} - these observations
have been primarily in the quasistatic regime, where the beads stay
in contact rather than undergoing collisions. Force measurements
using a photoelastic plate at the base of a sheared, cylindrical
pack of beads found a change in the distribution of forces as
jamming was approached \cite{corwin}. Earlier measurements in flow
in a rotating drum \cite{bonamy} found evidence of clustering, but
with a power law distribution of cluster sizes and hence no chosen
length scale. Growing spatial correlations were seen at the free
surface of chute flow down a plane \cite{pouliquen}, with a length
scale of the order of a few grain diameters. Previous simulations of
flow in a vertical tube geometry indicated an inhomogeneous
distribution of stresses \cite{denniston_li}. Earlier results
\cite{ally_epl04} on the same simulations we report on in this
article indicated that the most frequently colliding particles
organize into chain-like structures that form repeatedly and break
up as the particles move down the hopper. It was found that the
chain direction coincided with the principal axis of the collisional
stress in the system, with the lifetime of correlations in the
stress fluctuations increasing with decreasing flow rate
\cite{ally_pre}. There was quantitative agreement between the
simulational results and experiments in a two-dimensional hopper
geometry \cite{longhi,gardel}.

The absence of a clear structural signature of the approach to
jamming has led various groups to look for spatial inhomogeneities
in the dynamics. This approach was pioneered in structural glasses
where a variety of techniques probing local response functions
showed \cite{ediger} that the dynamics of a sample became
increasingly heterogeneous near the glass transition. Two point
correlation functions did not show clear signatures of
heterogeneity, however, an analysis of four-point correlation
functions in simulations of supercooled liquids
\cite{chandan,glotzer} showed evidence for growing spatial
correlations between localized density autocorrelations. Evidence
for dynamic heterogeneity was found in experiments on colloidal
glasses, in which highly mobile particles were found to cluster
\cite{weeks}, with a cluster size that increased as the glass
transition was approached. Heterogeneous dynamics have now also been
seen in experiments on dense granular material under shear
\cite{dauchot}. In our simulations as well, spatial heterogeneities
were found in the mean-squared displacements of particles when a
method \cite{harrowell} of maximizing the difference between highly
mobile and less mobile regions of the sample was used
\cite{ally_epl07}. While the size of the spatial heterogeneity
varied between five and six particle diameters as a function of flow
velocity, the "cage-size" or length scale over which the
heterogeneity was maximized did increase as the flow velocity
decreased towards jamming. However this increase in length scale was
typically smaller than a particle diameter, and the connection with
the collisional dynamics and force chains was not clear.

In the current paper, we analyze the development of spatial
correlations in both kinetic and dynamical variables in the flowing
state, and show that the extent of these increases as jamming is
approached. We would like to emphasize that the changing length
scale is seen in the two-point correlation functions of the velocity
and stress. We also draw qualitative connections between these
correlations and the chains of frequently colliding particles.

In the sections to follow, we first describe the simulation, and our
method of defining coarse-grained velocity and stress fields. We
then discuss our results for the time-averaged fields and the
temporal and spatial correlations in both velocity and stress
fluctuations, and conclude with a discussion of our results.

\section{Description of simulation}
The results we describe here are obtained from a two-dimensional
event-driven simulation of bidisperse hard disks falling under the
influence of gravity in a vertical hopper. We use the same particle
dynamics as Denniston and Li \cite{denniston_li}, and have described
our setup in some detail in an earlier paper \cite{ally_pre}. To
summarize: the inter-particle collisions are instantaneous and
inelastic, and there is no friction between the particles. As a
result, momentum transfer between colliding particles always occurs
along the vector separating their centers. The relative velocity
between colliding particles $i$ and $j$ is reduced by a coefficient
of restitution $\mu$, defined in the usual way:
\begin{equation}
(\bm{u}_{j}^{\prime} - \bm{u}_{i}^{\prime}).\hat{q} = -\mu (\bm{u}_j
- \bm{u}_i).\hat{q}
\end{equation}
where $\bm{u}_{j}^{\prime}$ and $\bm{u}_{i}^{\prime}$ are the
particle velocities after the collision, and $\hat{q}$ is a unit
vector along the line separating the centers of the particles.
Frictional effects at the wall are simulated by introducing a
coefficient of restitution $\mu_{\text{wall}}$ in the tangential
direction: the loss of vertical momentum at the walls is what allows
the flow to reach a steady state. In order to avoid inelastic
collapse, all collisions become elastic when the relative velocity
at the collision is below a certain threshold $u_{\text{cut}}$. A
particle exiting the base has a probability $p$ of being reflected,
else it exits the system and is re-introduced at the top. The flow
rate of particles in steady state is controlled by the size of the
opening at the base. The results described here are for a simulation
of 1000 particles of diameter 1 and 1.2 respectively, where grains
are chosen at random to have one or the other size. The other
simulation parameters are $\mu = 0.8$, $\mu_{\text{wall}} = 0.5$, $p
= 0.5$ and $u_{\text{cut}}= 10^{-3}$, and the mass of the smaller
grains is set to 1. Lengths are expressed in units of the smaller
particle diameter. In these units, the rectangular region of the
hopper has width 20 and height 76.5. The simulation is run for a
total of 1000 simulation time steps, of which the first 500 are
discarded. In units of the simulation time, the average time between
collisions for a given particle is on the order of $10^{-3}$.

Earlier results reported on these simulations \cite{ally_epl04,ally_pre}
were based on a particle-based analysis of the system as the size
of the opening, or the flow rate, was decreased.
Over a timescale larger than a typical collision time but shorter than the time
taken for a particle to fall through its own diameter,
particles with the highest frequency of collisions appear to repeatedly organize
into linear structures that form and break. These structures
were shown to carry much of the collisional stress \cite{ally_pre},
and their lifetime increased with decreasing flow rate, but no evidence
was found for a growing length scale.

In this paper, we seek to go beyond the particle-level analysis of
the system in order to quantify the correlations signaled by the
frequently colliding chains of particles and look for indications of
increasing order in the system as it approaches jamming. Thus in the
work described here, we have constructed coarse-grained variables,
looking at the system in terms of velocity, stress and density
fields and the spatial and temporal variations of these fields. We
also view this approach as a useful first step in developing a
continuum description of granular flow.

\section{Results}

\subsection{Coarse-grained Fields}
The system area is divided into square boxes of side equal to two
particle diameters. We define a box velocity \begin{equation}
\bm{v}(\bm{r},t) = \sum_{j} \bm{u}_j(t)/N_{j}
\end{equation} where $\bm{u}_j$ are the velocities of the $N_j$ particles
whose centers lie inside a box of center coordinates $\bm{r} =
(x,y)$ at a given instant $t$. We do this separately for the
vertical ($v_y$, parallel to flow) and horizontal ($v_x$,
perpendicular to the flow direction) velocity components.

\begin{figure}
\includegraphics{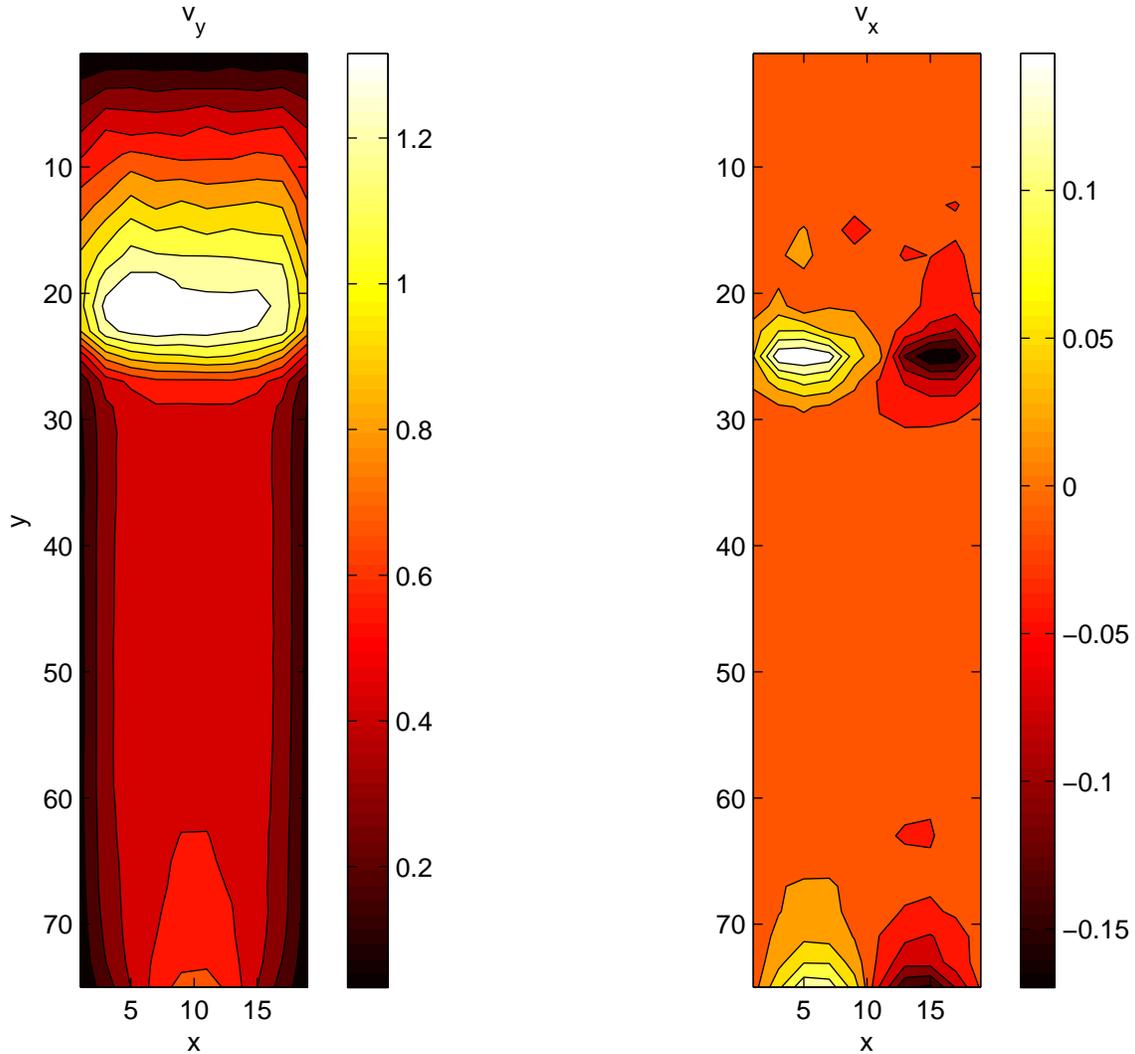}
\caption{\label{fig:vfield} (Color online) Contour plot of the
time-averaged spatial profile of the velocity components. (a) shows
the component parallel to the flow, and (b) shows the component of
velocity perpendicular to the flow. We focus on the region where the
velocity remains approximately constant.}
\end{figure}

Figure~(\ref{fig:vfield}) shows the time-averaged profile of the
velocity components for the slowest flow rate we report,
$v_{\text{flow}}$ = 0.60 in units of particle diameters per
simulation time. Figure~(\ref{fig:vfield}a) shows the vertical
velocity component $v_y$ over the entire hopper. There is an
acceleration region near the top of the hopper - particles are
introduced here, and accelerate under gravity before they reach an
asymptotic density and velocity. For all the analysis described in
the rest of the paper, we will focus on this region of constant
velocity, extending from y = 35 to y = 70. The correlation functions
we will present will also exclude the shear layer near the wall,
where the velocity is smaller, but non-zero.
Figure~(\ref{fig:vfield}b) shows the horizontal velocity component
field $v_x$ which is structureless in the region of constant
vertical velocity. In the acceleration region, the particles on
either side seem to be moving towards the center.

The time-averaged stress fields are calculated by a similar
coarse-graining procedure. The box collisional stress is defined as
follows: \begin{equation} \sigma^{\mu\nu}(\bm{r},t)= \frac{1}{\tau
A}\sum_{\text{collisions in } \tau} I^{\mu}r^{\nu} \end{equation}
where we sum over all collisions in a box occurring within the time
interval $[t,t+\tau]$. $\bm{I}$ is the impulse transferred at the
moment of collision, $\bm{r}$ the center to center vector between
colliding particles, and $A$ the box area. Since the particles are
hard disks with no inter-particle friction, all the momentum
transfer is in the direction of the vector separating the centers of
the two colliding particles. The above expression then gives us the
four stress components in the lab frame. We pick an averaging time
interval $\tau$ that is long compared to the typical collision time
but short compared to the time taken for a particle to fall through
its own diameter. These scales are well-separated in a dense flow:
both in experiments and in our simulation a particle undergoes many
collisions in the time it takes to fall through its own diameter.
Thus while the velocity profiles indicate that the central region of
the hopper is moving as a plug, the flow is highly collisional and
far from static, as found earlier in experiments \cite{menon} and
subsequently in simulations \cite{denniston_li}.

\begin{figure}
\includegraphics{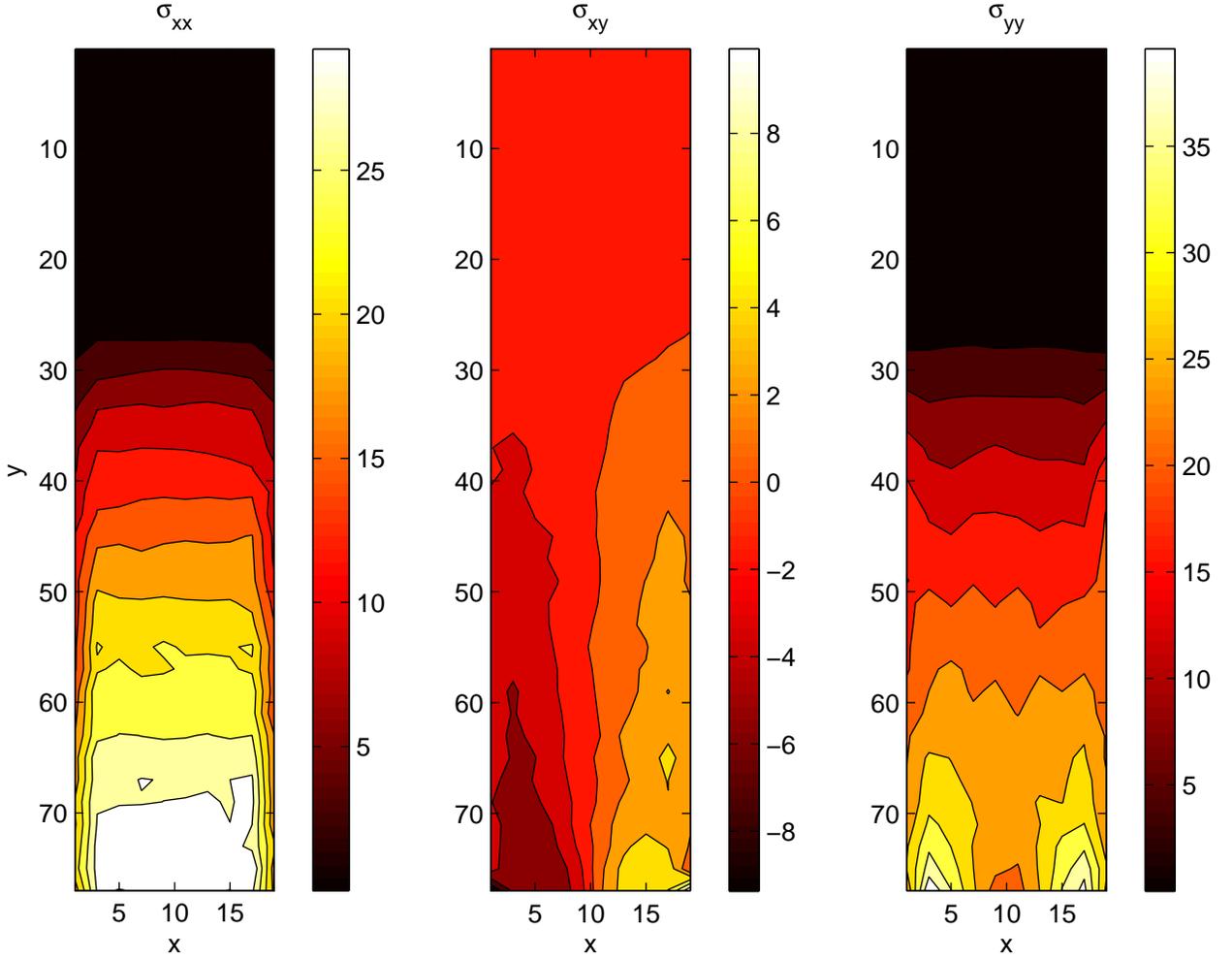}
\caption{\label{fig:sfield} (Color online) Contour plots (a) - (c)
of the time-averaged profiles of the stress tensor components
$\sigma_{xx}$ (horizontal pressure), $\sigma_{xy}$ (shear stress)
and $\sigma_{yy}$ (vertical pressure) respectively.}
\end{figure}

The time-averaged profiles of the stress components are shown in
Figure~(\ref{fig:sfield}). These are calculated as indicated in the
previous paragraph and averaged over the duration of the entire
simulation. Figure~(\ref{fig:sfield}a) and ~(\ref{fig:sfield}c) show
the normal stress components in the x and y directions respectively.
Note the similarity in structure in both, with a positive gradient
in both $\sigma_{xx}$ and $\sigma_{yy}$ in the direction of flow.
Thus there is an overall gradient in the pressure in the region of
constant average velocity. Since we only include the collisional
contributions to the stress, there is zero pressure in the
acceleration region at the top of the tube since there are no
collisions occurring in this region. Figure~(\ref{fig:sfield}b)
shows the shear stress component. The shear stress changes sign from
the left hand side to the right hand side of the tube, and has a
gradient in the horizontal direction. This appears consistent with
the formation of arch-like structures that span the tube, arising
from chains of frequently colliding particles that were reported
earlier \cite{ally_epl04}. Momentum transfer occurs along the
collisional chain and into the walls, but the direction of the wall
normal changes from one side of the tube to the other. The stress
tensor is symmetric on average $\sigma_{xy}=\sigma_{yx}$ since there
is no net torque on the system.

\begin{figure}
\includegraphics{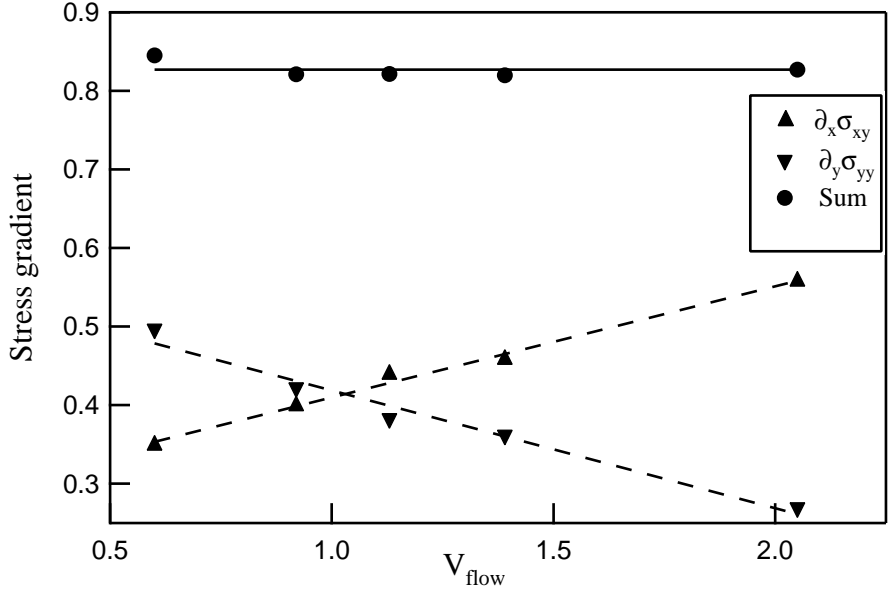}
\caption{\label{fig:sheargrad} A comparison of the spatial
derivatives of the stress components $\partial_x \sigma_{xy}$ and
$\partial_y \sigma_{yy}$ that support the weight of the column. Note
that their sum stays approximately constant as the flow rate
changes, but at high flow velocities the horizontal gradient of the
shear stress dominates, whereas the vertical gradient of the
vertical pressure becomes more important as the flow rate decreases.
}
\end{figure}

Momentum conservation dictates that
\begin{equation}\partial_t \rho v_i + \partial_j \rho v_i v_j =
\partial_j \sigma_{ij} + f_i \end{equation}
where $\rho$ is the local density, $v_i$ the velocity, $\sigma_{ij}$
the stress tensor and $f_i$ the external force. Under steady state
conditions, both terms on the left hand side are zero, and the
stress tensor on the right hand side is purely collisional. In
particular,
\begin{equation}\partial_x \sigma_{xy} + \partial_y \sigma_{yy} + f_y =
0 \end{equation} i.e. the sum of the vertical gradient in the
vertical pressure and the horizontal gradient in the shear stress
support the weight of the column. For our range of parameters, we
find that the relative importance of these two terms changes
smoothly with the flow rate: at high flow rates, the horizontal
gradient of the shear stress dominates, but at low flow rates, the
vertical gradient in the vertical pressure component is larger, as
shown in Figure~(\ref{fig:sheargrad}). Denniston and Li
\cite{denniston_li} reported that the horizontal gradient in the
shear stress gave a much larger contribution; but this was for a
single flow rate, which was larger than the largest flow rate we
study, and is therefore consistent with our findings.

\subsection{Relaxation Time: Velocity and Stress Autocorrelations}
We first examine fluctuations in the kinematic variables, and how
closely fluctuations at a given spatial point in the flow remain
correlated in time. The normalized autocorrelation $C_i(r,t)$ of the
velocity component $v_i(r,t)$ at a spatial point $r$ is defined as
\begin{equation} C_i(r,t) = \frac{\langle \Delta v_i(r, t+\tau)
\Delta v_i(r,t)\rangle}{\langle
   (\Delta v_i(r,t))^2\rangle} \label{eq:auto} \end{equation}
where $\Delta v_i(r,t) = v_i(r,t) - \langle v_i \rangle$ represents
the time-dependent fluctuation of the velocity component, $i = x$ or
$y$, and the averages are over time. This quantity is then spatially
averaged over all the boxes in the region of constant velocity,
giving
\begin{equation} \langle v_i(t) v_i(0) \rangle = \frac{1}{N_s}\sum_r
C_i(r,t)\label{eq:autosum} \end{equation} where $N_s$ is the number
of boxes summed over.

\begin{figure}
\includegraphics{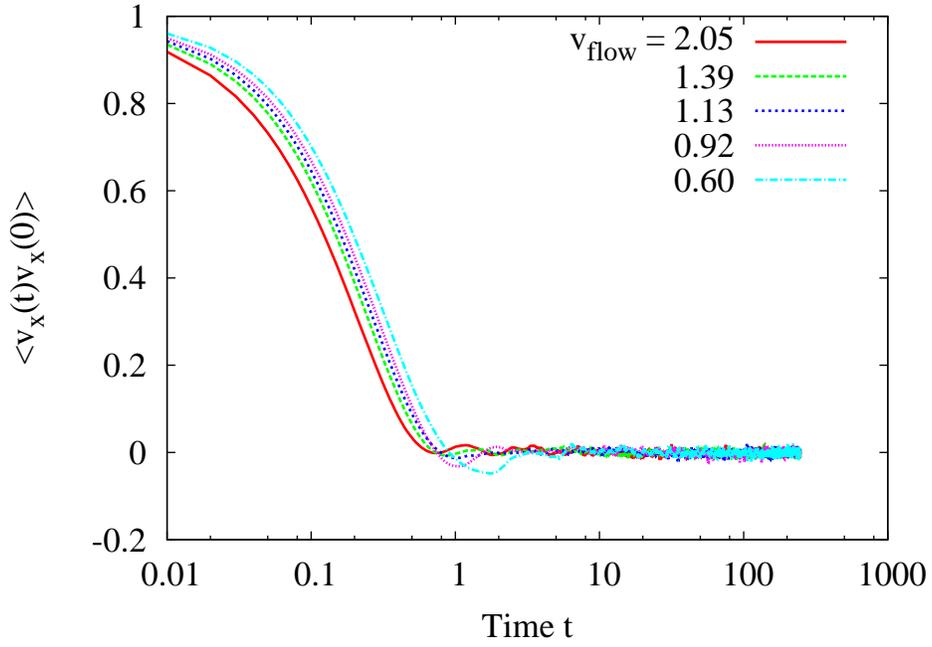}
\caption{\label{fig:vxauto} (Color online) Autocorrelation of the
velocity component perpendicular to the flow at at five different
flow rates. The flow rates are expressed in units of particle
diameters per simulation time.}
\end{figure}

\begin{figure}
\includegraphics{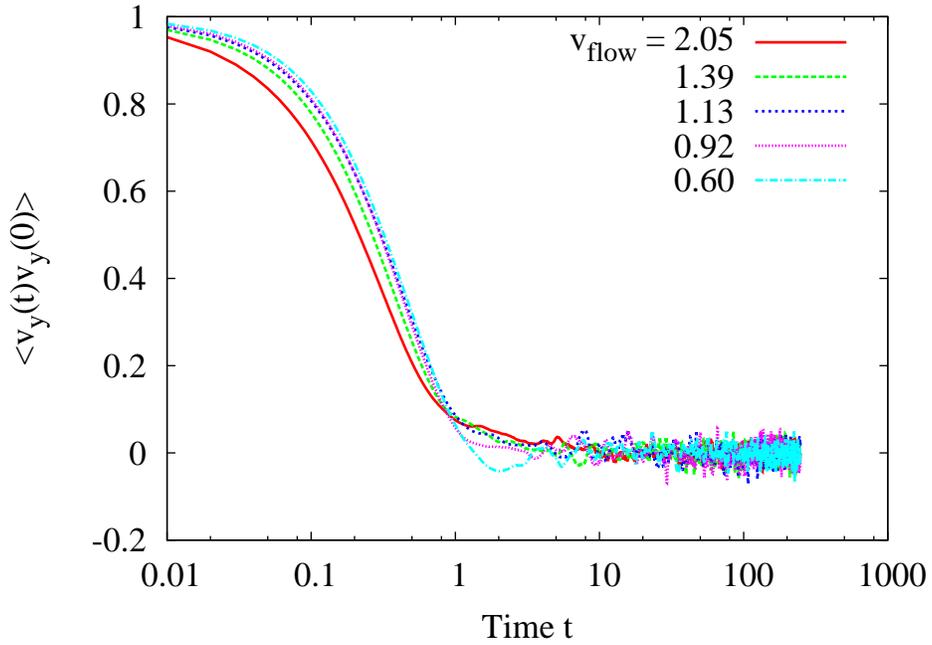}
\caption{\label{fig:vyauto} (Color online) Autocorrelation of the
velocity component parallel to the flow at the same five flow
rates.}
\end{figure}

The autocorrelations of the fluctuations in both velocity components
are shown as a function of time in Figures~(\ref{fig:vxauto}) and
(\ref{fig:vyauto}) for five different flow velocities. The
correlations fall off fairly rapidly at short times at all flow
rates. As one moves from fast to slow flow (from left to right in
the figures), there is a consistent but slow trend towards
increasing relaxation time as the flow rate decreases (note that the
x-axis scale has been made logarithmic in the figure to emphasize
this). Thus temporal correlations in the velocity fluctuations do
not provide a strong indication of an impending jam as the flow rate
decreases.

\begin{figure}
\includegraphics{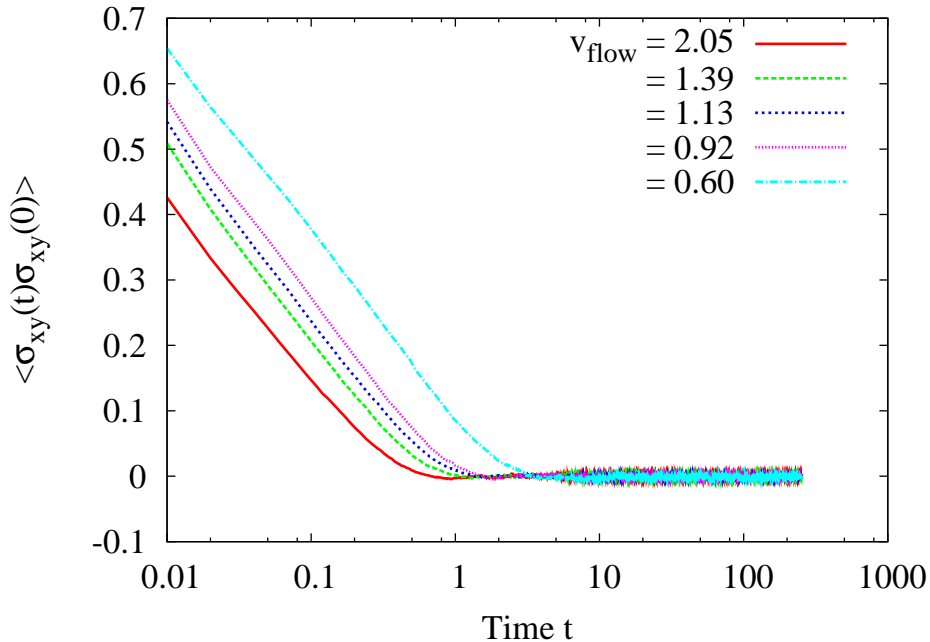}
\caption{\label{fig:sxyauto} (Color online) Autocorrelation of the
shear stress at the same five flow rates.}
\end{figure}

Clearer evidence of a growth in the relaxation timescale is seen in
the stress. The autocorrelation of the fluctuations in the stress
components $\langle \sigma_{ij}(t) \sigma_{ij}(0) \rangle$ is
defined exactly the same way as for the velocity components, as
indicated in Eqs.~(\ref{eq:auto}) and (\ref{eq:autosum}). These
autocorrelations decay with time with some characteristic timescale;
and evidence for an increase in decay timescale as the flow-rate
decreases is seen in all three components. However, the increase in
timescale is most pronounced for the shear stress $\sigma_{xy}$, and
the autocorrelation of this stress component is shown in
Figure~(\ref{fig:sxyauto}) for five flow rates. Note that the time
axis in this figure is logarithmic, and after an initial drop at
very short times (equivalent to one averaging timestep), the decay
of the stress autocorrelations is logarithmic at intermediate times.
The change with flow rate of the autocorrelation time (measured as
described in the next paragraph) of the two other stress components,
the horizontal and vertical pressures $\sigma_{xx}$ and
$\sigma_{yy}$ respectively, is similar to that seen for the two
velocity components in Figs.~(\ref{fig:vxauto}) and
(\ref{fig:vyauto}).

\begin{figure}
\includegraphics{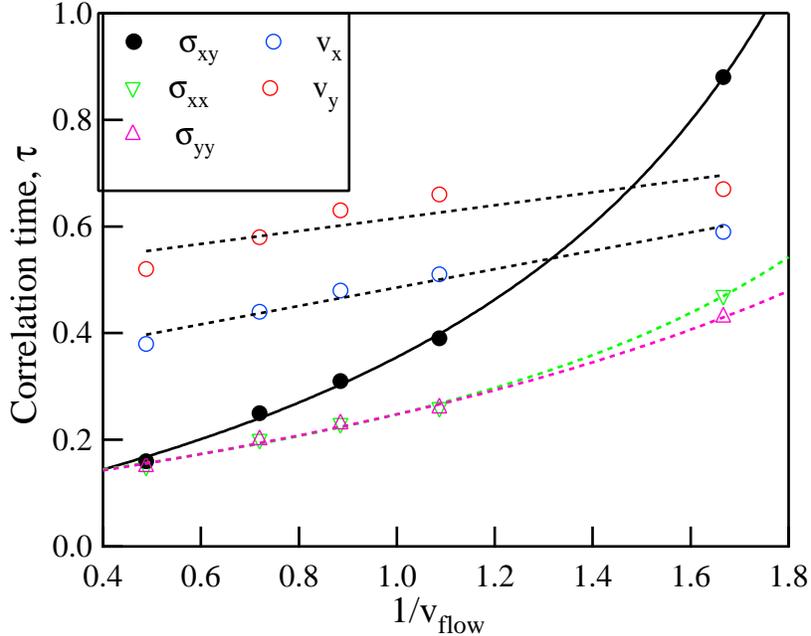}
\caption{\label{fig:tau} (Color online) Time at which the
autocorrelations of the different velocity and stress components
drop to 0.1 as a function of the inverse flow velocity. The lines
are intended as guides to the eye and are obtained from quadratic
fits to the data. Note the timescale associated with the shear
stress shows a much greater dependence on flow velocity.}
\end{figure}

We extract a time scale from these autocorrelation functions by
measuring the time at which the (normalized) autocorrelation
function drops to 0.1. This time $\tau$ is plotted as a function of
inverse mean flow velocity in Figure~(\ref{fig:tau}). Note that
while the autocorrelation time associated with the two velocity
components and the two diagonal components of stress all increase as
flow rate decreases, the timescale associated with the shear stress
increases more rapidly than the others as the flow slows. This
increase is also more rapid than linear (the lines in the figure,
intended as guides to the eye, are fits to quadratics). As we shall
see in the next section, this increase in autocorrelation time leads
to the growth of a region that flows like a plug, and is consistent
with the presence of chains of frequently colliding particles that
begin to span the system at the lower flow rates. Evidence that the
dominant contribution to the principal axis of the collisional
stress tensor comes from the most frequently colliding particles was
presented in earlier work \cite{ally_pre} on the same simulations.
There we found that the timescale associated with the fluctuations
of the principal axis of the stress increased linearly with inverse
flow rate.

\subsection{Length Scales: Spatial correlations in velocity and stress}
We next examine the spatial correlations in the fluctuations of both
kinematic and dynamic variables. We define the normalized equal time
spatial correlation function of the velocity fluctuations as a
function of separation as follows:
\begin{equation}\langle v_i(\bm{r}) v_i(0) \rangle =
\frac{\sum_{\bm{r}_i}\sum_{t}\Delta v_i(\bm{r}_i, t) \Delta v_i((\bm{r}_i
+ \Delta \bm{r}, t)}{N_s \sum_t (\Delta v_i(\bm{r}_i, t) )^2}
\label{eq:space}
\end{equation} where $N_s$ is the number of spatial points (boxes)
summed over, and $\Delta v_i(r,t)$ represents the velocity
fluctuation relative to the long-time average.

\begin{figure}
\includegraphics{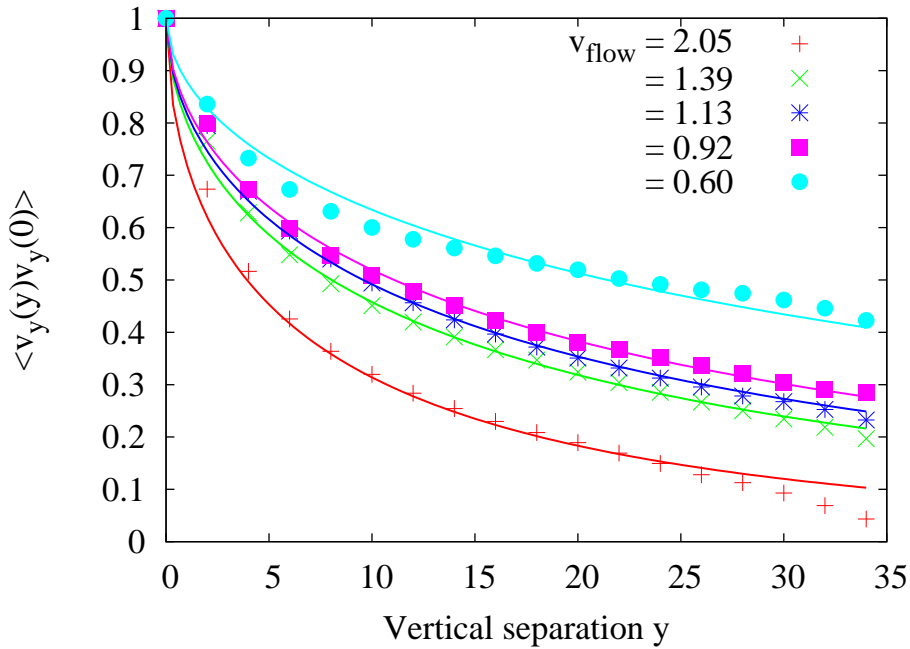}
\caption{\label{fig:vyspace} (Color online) The equal time spatial
correlation function of the vertical velocity component is shown as
a function of vertical separation at five different flow rates. The
points are obtained from the numerical simulation, and the solid
lines represent fits to the data using the stretched exponential
function $\exp[-(x/\lambda)^{0.55}]$ with $\lambda$ as the only
fitting parameter. }
\end{figure}

The spatial correlations for the vertical velocity component
$\langle v_y(y) v_y(0) \rangle$ are shown as a function of vertical
separation $y$ in Figure~(\ref{fig:vyspace}) for five different flow
rates. Note that as the flow rate decreases, the spatial correlation
dies off more slowly, suggestive of an increasing length scale. Our
numerical results are shown by the points in the figure, and the
solid lines are fits to stretched exponential behavior - we shall
discuss these fits further in what follows. The figure shows that
the correlations are very long range and complete decorrelation is
not achieved across the entire region of analysis. We are beginning
to systematically investigate finite size effects on the growing
length and time scales in order to better characterize the nature of
the jamming transition.

\begin{figure}
\includegraphics{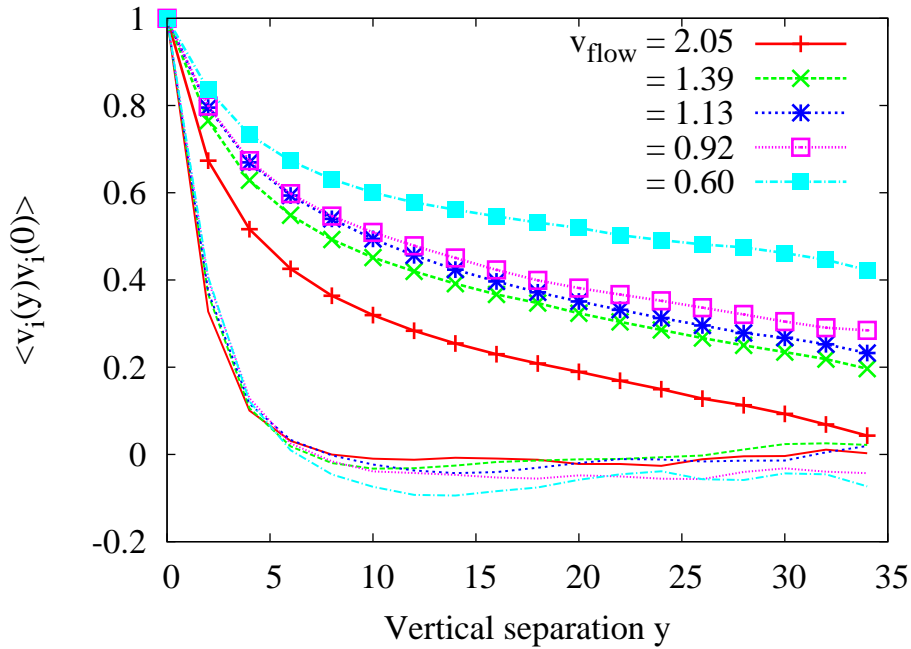}
\caption{\label{fig:vxvyspace} (Color online) Comparison of the
equal time spatial correlation function of the vertical velocity
component from the previous graph (lines and symbols) to the spatial
correlations of the horizontal velocity component (lines) as a
function of vertical separation at the same five flow rates. No fits
are shown in this plot, and the lines simply connect the points.}
\end{figure}

The spatial correlation function for the horizontal velocity
component $v_x(y)$ is shown in Figure~(\ref{fig:vxvyspace}), along
with the curves from the previous figure using the same colour
scheme for the same set of flow rates. Figure~(\ref{fig:vxvyspace})
shows that spatial correlations in the horizontal velocity component
die out very rapidly, and there is no systematic change in the
spatial extent of the correlations with flow rate. The same trend is
seen when the spatial correlations of both velocity components are
computed as a function of horizontal separation $\langle v_i(x)
v_i(0) \rangle$.

\begin{figure}
\includegraphics{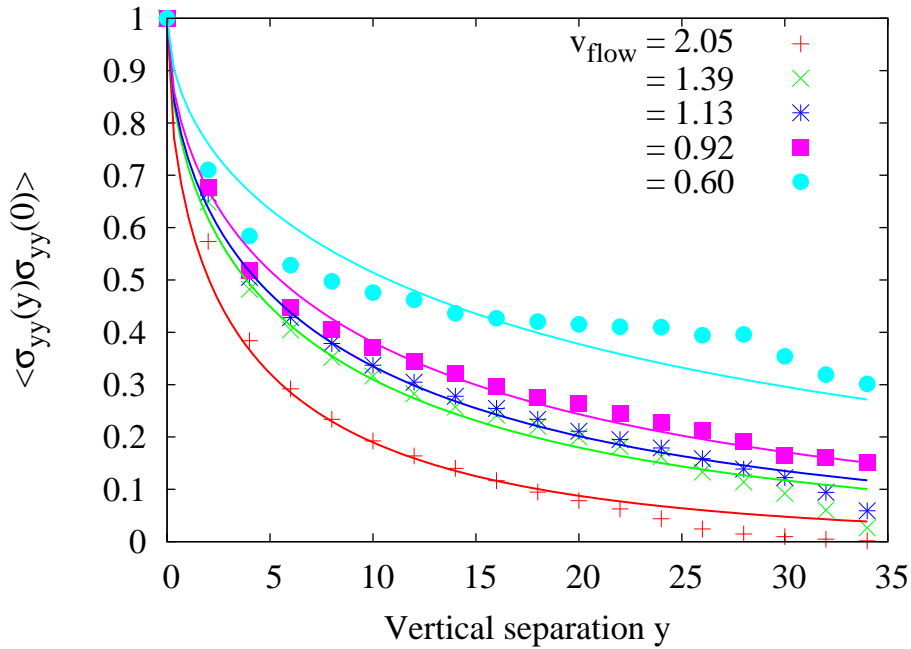}
\caption{\label{fig:syyspace} (Color online) Plots of the equal time
spatial correlation functions of the vertical pressure fluctuations
$\Delta \sigma_{yy}$ as a function of vertical separation at five
flow rates. The points represent the numerical data, and the solid
lines are once again fits to stretched exponential behavior
$\exp[-(x/\lambda)^{0.55}]$.}
\end{figure}

We next compute the equal-time spatial correlation function for the
components of the stress tensor exactly as in Eq.~(\ref{eq:space}).
The spatial correlations for the vertical pressure $\langle
\sigma_{yy}(y) \sigma_{yy}(0) \rangle$ are shown as a function of
vertical separation $y$ in Figure~(\ref{fig:syyspace}) for five
different flow rates. This looks very similar to the behavior seen
in Figure~(\ref{fig:vyspace}) for the spatial correlations in the
vertical velocity. Once again, the points are obtained from the
simulations and the lines represent fits to stretched exponential
behavior.

Spatial correlations in the horizontal pressure $\langle
\sigma_{xx}(x) \sigma_{xx}(0) \rangle$ as a function of horizontal
separation show very similar behavior to that seen in
Figure~(\ref{fig:syyspace}), but we have not shown it because of the
limited range available in the $x$-direction. Equal-time spatial
correlations of the vertical pressure $\sigma_{yy}$ as a function of
horizontal separation $x$ and those of the horizontal pressure
$\sigma_{xx}$ as a function of vertical separation $y$ (both not
shown) also have the same trend as a function of flow rate, but show
a less pronounced change than the correlations seen in
Figure~(\ref{fig:syyspace}). Interestingly, the shear stress shows
no spatial correlations: the equal-time spatial correlation function
in $\sigma_{xy}$ dies very rapidly and shows no change as a function
of flow rate. We will return to this observation in the next
section.

\begin{figure}
\includegraphics{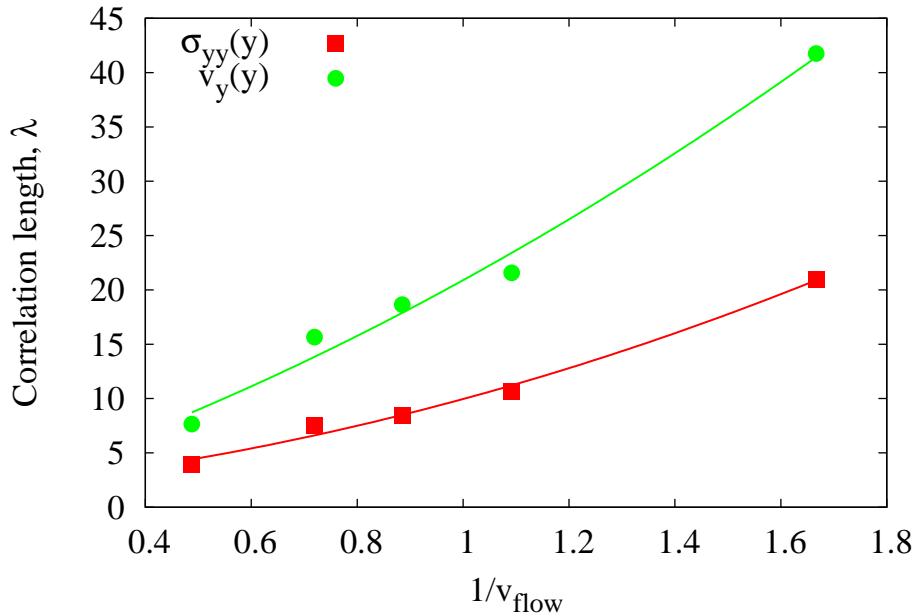}
\caption{\label{fig:lambda} (Color online) The length scales
$\lambda$ obtained from fitting the equal time spatial correlation
functions $\langle v_y(y) v_y(0) \rangle$ and $\langle
\sigma_{yy}(y) \sigma_{yy}(0) \rangle$ to a stretched exponential
decay $\exp[-(x/\lambda)^{0.55}]$ are plotted as a function of
inverse flow rate.}
\end{figure}

We can extract from Figures~(\ref{fig:vyspace}) and
(\ref{fig:syyspace}) a characteristic length scale $\lambda$ by
fitting both spatial correlation functions to a stretched
exponential decay function, $\exp[-(x/\lambda)^{0.55}]$ where
$\lambda$ is the only fitting parameter. (We estimated the exponent
$0.55$ by first doing a two-parameter fit of the curves.) The fits
are represented by solid lines in the two figures. We plot $\lambda$
as a function of inverse flow rate in Figure~(\ref{fig:lambda}).
Though the $\lambda$ values obtained from $v_y$ and $\sigma_{yy}$
are not the same, both increase faster than linearly with inverse
flow rate and follow the same trend.

\begin{figure}
\includegraphics{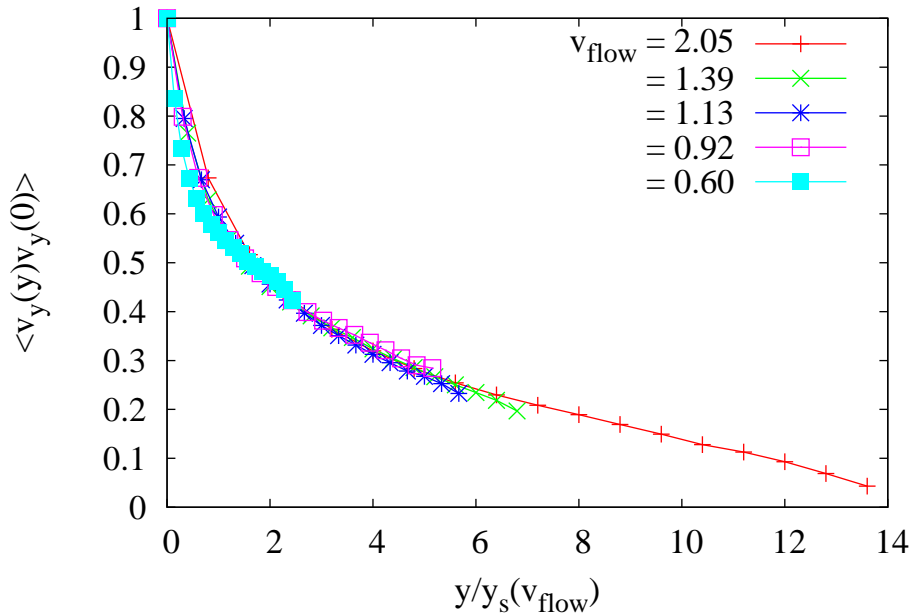}
\caption{\label{fig:rescaled} (Color online) The equal time spatial
correlation function of the vertical velocity component $\langle
v_y(y) v_y(0) \rangle$ is plotted as a function of scaled vertical
position $y/y_s(v_{flow})$ where $y_s(v_{flow})$ is a flow-velocity
dependent length. The curves do not superimpose exactly, with a
significant change in shape as the flow approaches jamming.}
\end{figure}

It is clear from Figures~(\ref{fig:vyspace}) and
(\ref{fig:syyspace}) that the correlation functions are not
perfectly described by stretched exponential behavior: the fits work
best at intermediate flow velocities, but there are deviations at
slow and fast flow, and the agreement is very poor at the slowest
flow rate. We tried other ways of extracting a length: by measuring
the distance at which the correlation function drops to a fixed
fraction, say 0.4; and by scaling the horizontal axis by a
flow-velocity dependent length to superimpose all the curves. In all
these cases, though the actual $\lambda$ values change, the
dependence of the length on flow velocity shows the same trend as in
Figure~(\ref{fig:lambda}). Rescaling the horizontal axis also shows
us that the shape of the correlation function changes as one
approaches jamming, see Figure~(\ref{fig:rescaled}).

\begin{figure}
\includegraphics{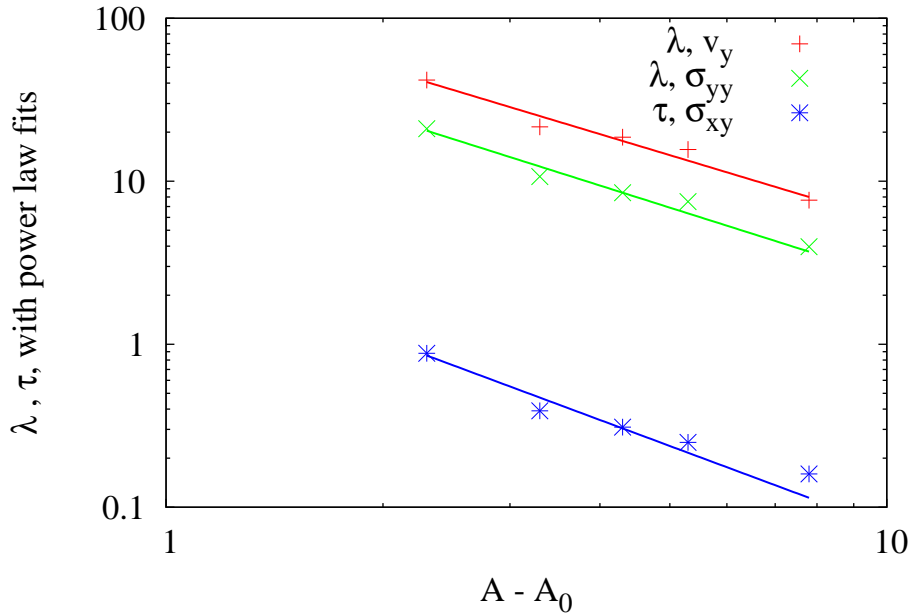}
\caption{\label{fig:tau_lambda} (Color online) The length scales
$\lambda$ obtained from the equal time spatial correlation functions
of $v_y$ and $\sigma_{yy}$ and the timescale $\tau$ from the
autocorrelation of $\sigma_{xy}$ are plotted as a function $A -A_0$,
where $A$ is the size of the opening, and $A_0$ the opening at which
the flow velocity extrapolates to zero. The solid lines are power
law fits with exponents -1.3 ($\lambda,v_y$), -1.4 ($\lambda,
\sigma_{yy}$), and -1.6 ($\tau,\sigma_{xy}$).}
\end{figure}

Finally, the length scales $\lambda$ extracted from the vertical
velocity $v_y$ and pressure $\sigma_{yy}$ fluctuations
(Fig.~\ref{fig:lambda}) as well as the timescale $\tau$ obtained
from the autocorrelations of the shear stress $\sigma_{xy}$
(Fig.~\ref{fig:tau}) are plotted as a function of $A -A_0$ in
Figure~(\ref{fig:tau_lambda}). Here $A$ is the size of the opening
at the bottom of the hopper, and $A_0=2.2$ is the size at which the
flow velocity extrapolates to zero. (The flow velocity decreases
roughly linearly with the size of the opening). All three quantities
appear to display power law behavior as indicated by the solid line
fits, though our dynamic range is too small to say this more
definitively. The power law exponents for the length scale in the
vertical velocity fluctuations is -1.3, close to the -1.4 exponent
seen for the length scale coming from the vertical pressure
fluctuations. The exponent for the timescale in the shear stress
fluctuations is -1.6. It is difficult to make measurements over a
larger range of flow velocities for this particular system, since a
steady state is hard to achieve when the opening is too wide, and
the flow tends to jam intermittently when the opening is less than
four particle diameters. We plan to run the simulation for larger
systems which would help us better measure these power laws.

\section{Discussion}
We have observed a growing length scale in dense gravity-driven
granular flow as the flow rate decreases towards jamming. This
length scale characterizes the decay of the two-point spatial
correlation functions of the velocity and the normal stress.
Correspondingly, there is a growth in the relaxation time associated
with the shear stress. By contrast, the relaxation times of the
velocity fluctuations grow very little, and no accompanying
structural changes in the density are seen. Thus increasing temporal
correlations in the shear stress fluctuations are closely tracked by
spatial correlations in the flow velocity and the pressure. Both the
relaxation times and length scales derived from these two-point
correlation functions appear to increase as power laws of the
opening size in the dynamic range available to us.

An interesting scenario, that we are in the process of
investigating, is whether an increasing length scale in two-point
spatial correlation functions of the velocity, as seen here, can
lead to the type of behavior observed in four point density
correlation functions \cite{dauchot,keys}.

In standard critical phenomena it is usually a diverging length
scale that signals the development of order in the system, and leads
to a divergence in the relaxation times. Here by contrast, our
observations indicate that it is the timescale associated with the
collisional dynamics that leads to an increase in length scale in
the stress. A connection between the structure of the stress tensor
and frequently colliding chains of particles was previously
established \cite{ally_pre}. The increase in autocorrelation time of
the shear stress as the flow rate decreases indicates an increase in
lifetime of these collision chains. These chains form and break up
repeatedly during the flow, and while they lead to an increased
autocorrelation time in the \emph{shear stress}, no corresponding
increase in the spatial correlation of the shear stress fluctuations
is seen. The effect of their formation is instead to make the system
flow as a plug, which is seen in increased spatial correlations of
the \emph{normal} stresses and vertical velocity. The increase in
the equal time spatial correlations of the velocity can be construed
as a further sign that collisions in the system are becoming less
random. The time between collisions decreases along with the flow
rate, so there are many more collisions in a given time interval.
Thus collisions must occur in a spatially correlated way.

Our results agree well with experiments on gravity-driven hopper
flow in two dimensions \cite{gardel} in which force-velocity
correlations have long-range effects. Our simulations make it clear
that these correlations are also present in the spatial structure of
the stress fluctuations. Our observation that an increasing length
is associated with the normal components of the stress and not in
the shear component poses a further puzzle.

\begin{acknowledgments}
We wish to acknowledge useful discussions with Narayanan Menon,
Nalini Easwar, Giulio Biroli, Andrea Liu and Douglas Durian.
We thank Melanie Finn and Anna-Lisa Baksmaty,
current and former undergraduates at Mount Holyoke College, for their
contributions to this project. ST and BC thank the Aspen
Center for Physics for their hospitality. BC acknowledges the support of
NSF-DMR 0549762.
\end{acknowledgments}

\end{document}